# Robustness of ferromagnetism in (In,Fe)Sb diluted magnetic semiconductor to variation of charge carrier concentration and Fermi level position

A.V. Kudrin[1,*], V.P. Lesnikov[1], Yu.A. Danilov[1], M.V. Dorokhin[1], O.V. Vikhrova[1], I.N. Antonov[1], R.N. Kriukov[1], S.Yu. Zubkov[1], D.E. Nikolichev[1], A.A. Konakov[1], Yu.A. Dudin[1], Yu.M. Kuznetsov[1], N.A. Sobolev[2,3], and M.P. Temiryazeva[4]

[1]*Lobachevsky State University of Nizhny Novgorod, Gagarin av. 23/3, 603950 Nizhny Novgorod, Russia*
[2]*Department of Physics and I3N, University of Aveiro, 3810-193 Aveiro, Portugal*
[3]*National University of Science and Technology "MISiS", 119049 Moscow, Russia*
[4] *Kotel'nikov Institute of Radioengineering and Electronics of RAS, Fryazino Branch, 141190 Fryazino, Russia*
*kudrin@nifti.unn.ru

The influence of $He^+$ ion irradiation on the transport and magnetic properties of epitaxial layers of a diluted magnetic semiconductor (DMS) (In,Fe)Sb, a two-phase (In,Fe)Sb composite and a nominally undoped InSb semiconductor has been investigated. In all layers, a conductivity type conversion from the initial *n*-type to the *p*-type has been found. The ion fluence at which the conversion occurs depends on the Fe concentration in the InSb matrix. Magnetotransport properties of the two-phase (In,Fe)Sb layer are strongly affected by ferromagnetic Fe inclusions. An influence of the number of electrically active radiation defects on the magnetic properties of the single-phase $In_{0.75}Fe_{0.25}Sb$ DMS has been found. At the same time, the results show that the magnetic properties of the $In_{0.75}Fe_{0.25}Sb$ DMS are quite resistant to significant changes of the charge carrier concentration and the Fermi level position. The results confirm a weak interrelation between the ferromagnetism and the charge carrier concentration in (In,Fe)Sb.

## I. INTRODUCTION

Diluted magnetic semiconductors (DMS) have attracted great interest because of their potential for semiconductor spintronics [1]. During the last two decades many important results were obtained for III-V semiconductors heavily doped with Mn, in particular for (Ga,Mn)As. However, the Curie temperature ($T_C$) of Mn doped semiconductors is relatively low (up to ~ 190 K for GaMnAs [1]), which limits their possible application. Now III-V semiconductor layers heavily doped with Fe are new interesting materials for the semiconductor spintronics. In particular, (Ga,Fe)Sb [2,3] and (In,Fe)Sb [4,5] layers with a room temperature (RT) Curie point were obtained. At present, the nature of ferromagnetism in Fe doped III-V semiconductors is a subject of research. In the case of (In,Fe)As, a carrier-mediated mechanism was suggested [6]. For (Al,Fe)Sb [7], (Ga,Fe)Sb and (In,Fe)Sb the ferromagnetism seems to be associated with some superexchange interaction between Fe atoms without the determinative role of the charge carrier concentration.

Irradiation of semiconductors with accelerated light particles leads to the creation of electrically active radiation defects (RDs), in particular vacancies and antisites. In the GaAs matrix RDs form deep donor and acceptor levels near the middle of a band gap, that usually reduce the carrier concentration [8.9]. The creation of RDs in (Ga,Mn)As makes it possible to vary the carrier density and to analyze the carrier density influence on the transport and magnetic properties [10-12]. The consequence of RDs creation in the GaSb and InAs matrices is the appearance of additional charge carriers (holes in GaSb and electrons in InAs), which is accompanied by a Fermi level ($E_F$) shift into the valence or conduction band in GaSb [13] and InAs [14], respectively. In the InSb matrix the radiation defects manifest themselves in a more complex way - both acceptor and donor centers appear [15,16]. Thus, the irradiation of (Ga,Fe)Sb, (In,Fe)Sb and (In,Fe)As diluted magnetic semiconductors (DMS) with energetic light ions makes it possible to change the charge carrier density keeping a fixed concentration of introduced Fe atoms. As a consequence, a study of the influence of the carrier concentration and the Fermi level position on the magnetic properties of the Fe doped narrow-bandgap III-V semiconductors allows to clarify the nature of the observed ferromagnetism.

In this study, we present results of the charge carrier density control in InSb and (In,Fe)Sb layers through the creation of RDs by means of helium ion irradiation with fluences in the range from $1 \times 10^{13}$ – $1 \times 10^{16}$ cm$^{-2}$.

## II. EXPERIMENTAL

The InSb and (In,Fe)Sb layers were grown by pulsed laser deposition in a vacuum on semi-insulating (001) GaAs substrates [4].The Fe content was set by the technological parameter $Y_{Fe} = t_{Fe}/(t_{Fe}+t_{InSb}+ t_{Sb}) = 0.25$, where $t_{Fe}$, $t_{InSb}$ and $t_{Sb}$ are the ablation times of the Fe, InSb and additional Sb targets [4], respectively.

The surface of the structures was examined by atomic force (AFM), magnetic force (MFM) and Kelvin probe force microscopy. Optical reflectivity spectra were obtained at RT in the spectral range from 1.6 – 6 eV. The elemental composition was determined by energy-dispersive X-ray photoelectron spectroscopy (XPS). The *dc* magnetotransport measurements were carried out in the van der Pauw geometry in a closed-cycle He cryostat. The layers were irradiated at RT with 50 keV He$^+$ ions with fluences ($F$) ranging from $1 \times 10^{13} - 1 \times 10^{16}$ cm$^{-2}$. To prevent channeling of the implanted ions, the ion beam was directed at an angle of about 10º off the normal to the (001) wafer surface and at an azimuthal angle of 45º.

Our previous studies revealed that the growth temperature ($T_g$) is a critical parameter for the (In,Fe)Sb layers [17]. A transmission electron microscopy and an energy-dispersive x-ray spectroscopy revealed a strong dependence of the phase composition of the (In,Fe)Sb compound on $T_g$ [17]. The (In,Fe)Sb layer with a Fe content of 10 at. % and $T_g$ = 300 ºC contained secondary crystalline phase inclusions (Fe clusters with a size of ~ 20 nm) formed due to the coalescence of Fe atoms. At the same time, the In$_{0.8}$Fe$_{0.2}$Sb layer with $T_g$ = 200 ºC was a single phase one with a relatively uniform distribution of Fe atoms [17].

The present paper contains results obtained on the following structures: an undoped InSb layer grown at 250ºC (sample 250-0, grown by sputtering an InSb target only), and two (In,Fe)Sb layers with $Y_{Fe}$ = 0.25 grown at 200 and 300ºC (samples 200-25 and 300-25). The thickness of the layers is about 80 nm. In our previous work [4] the structure fabricated using an additional Sb target had a corresponding symbol in the name (200(Sb)-17). In this paper we omit the symbol Sb in the names of the Fe doped structures.

## III. RESULTS

### A. Structural and optical properties

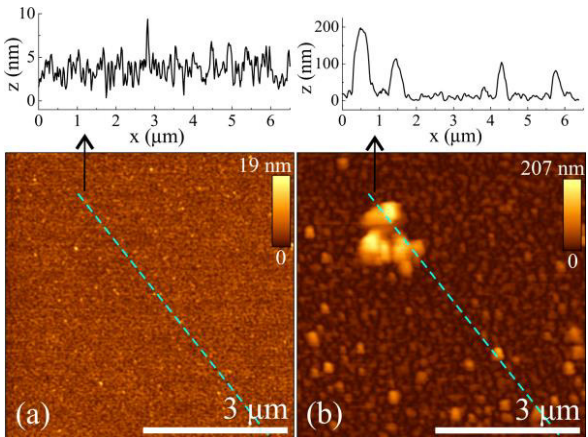

FIG. 1. AFM images and height profiles. (a) Structure 200-25. (b) Structure 300-25.

Figure 1 shows the AFM surface morphology of structures 200-25 (Figure 1(a) and 300-25 (Figure 1(b)). The (In,Fe)Sb layer grown at $T_g$ = 200ºC has a smooth surface with a root mean square (RMS) roughness of about 1.4 nm (Figure 1(a)). The surface of the (In,Fe)Sb layer grown at $T_g$ = 300ºC is much more rough (RMS ≈ 25 nm) with an array of islands (base diameter ~ 1 µm, height ~ 100 nm, Figure 1(b)).

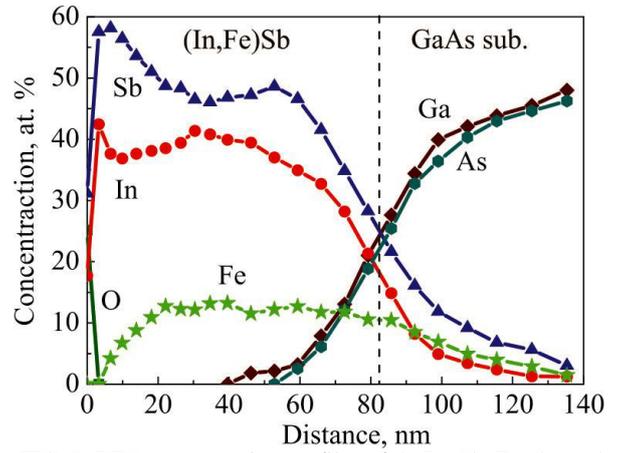

FIG. 2. XPS concentration profiles of O, In, Sb, Fe, Ga and As atoms for structure 200-25.

Figure 2 shows XPS dependences of the concentration of O, In, Sb, Fe, Ga and As atoms on the distance from the surface for sample 200-25. The average Fe content in the (In,Fe)Sb layer detected by XPS equals 12.5 ± 1 at. %. In Ref. [4] we previously obtained single-phase (In,Fe)Sb layers with a Fe content of 13 at. % and a Curie point above RT. However, during the growth of the (In,Fe)Sb layers described in Ref. [4], no additional amount of Sb was introduced (except for the structure 200(Sb)-17) and the formation of In-enriched islands on the surface was observed. In present study, the (In,Fe)Sb layer of structure 200-25 has a smooth surface (Figure 1(a)) and contains a similar amount of Fe (≈ 12.5 at. %), therefore, the Curie temperature is also expected to be close to RT.

For sample 300-25, the average Fe content in the (In,Fe)Sb layer is the same (inasmuch as the technological parameter $Y_{Fe}$ also is equal to 0.25), however, the distribution of Fe atoms is different as a result of the significantly higher growth temperature. Similar to what was observed for the (In,Fe)Sb layer with the Fe content of 10 at. % [17], the $T_g$ increase from 200 to 300ºC for sample 300-25 should lead to a coalescence of Fe atoms and to the formation of a secondary crystalline phase in form of Fe clusters within the (In,Fe)Sb layer.

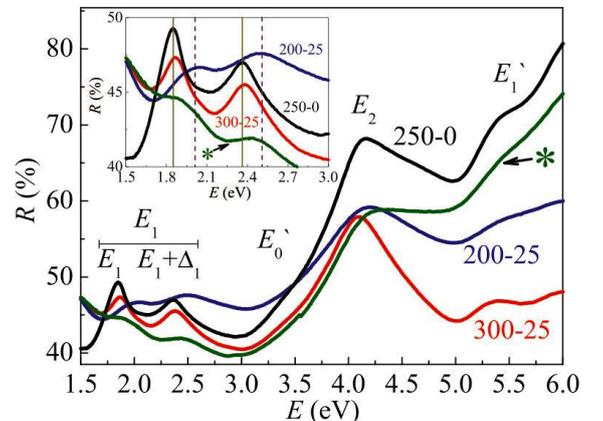

FIG. 3. Optical reflectivity spectra at 295 K of structures 200-0, 200-25 and 300-25. Symbol * corresponds to the spectrum of structure 200(Sb)-17 from Ref. [4]. The inset shows enlarged spectrum parts in the $E_1$ region.

Figure 3 shows reflectivity spectra measured at 295 K of structures 200-0, 200-25, 300-25 and of structure 200(Sb)-17 from Ref. [4] for comparison. The reflectivity spectra coincide with the spectrum of an InSb crystal and contain features associated with characteristic interband transitions [18]. In particular, the doublet in the $E_1$ region and the intense peak in the $E_2$ region are well resolved. The reflectivity spectrum for sample 300-25 is quite similar to the spectrum of the undoped InSb layer (sample 250-0). The peaks in the $E_1$ region are pronounced but have a slight blueshift. For sample 200-25 the $E_1$ peaks are less pronounced, and the blueshift is larger (see the inset to Figure 3). The spectrum of sample 200(Sb)-17 from Ref. [4] is similar to the spectrum of sample 200-25, but the blueshift is smaller than for sample 200-25, since the Fe concentration is lower. A linear blueshift of the $E_1$ peak position with the Fe concentration for (In,Fe)Sb layers grown by molecular beam epitaxy was observed in Ref. [5]. Using the linear blueshift dependence and the $E_1$ peaks positions for sample 250-0 and 200-25 (with a Fe content of about 12.5 at. %), the Fe concentration in the $In_{1-x}Fe_xSb$ matrix for sample 300-25 can be roughly estimated to equal ~ 1.5 at. % (i.e. $x$ ~ 0.03). Hence, the main fraction of the Fe atoms in sample 300-25 is in the form of second-phase iron inclusions. Based on the described results and our previous investigations [4,17], we may conclude, that the $In_{0.75}Fe_{0.25}Sb$ layer of the sample 200-25 is a single-phase DMS while the (In,Fe)Sb layer of sample 300-25 is the two-phase system – viz. a $In_{1-x}Fe_xSb$ ($x$ ~ 0.03) matrix with second-phase Fe inclusions.

### B. Transport and magnetic properties

Let us consider the influence of $He^+$ ions irradiation on the transport and magnetic properties of the structures. As mentioned in the Introduction, RDs in the InSb matrix can be both acceptors or donors. The as-grown InSb and (In,Fe)Sb layers are $n$-type due to the presence of native electrically active donor defects [4, 19]. After the irradiation with a fluence of $1 \times 10^{14}$ cm$^{-2}$ the InSb layer (sample 250-0) demonstrates a conversion from the $n$- to the $p$-type. Table 1 presents the experimental values of carrier concentration, type of conductivity and Hall mobility at 295 K for the as-grown sample 250-0 and and for that after irradiation with different fluences. The conductivity type conversion was confirmed by the Seebeck effect measurements at RT. For our InSb layer the concentration of acceptor RDs exceeds the concentration of donor radiation defects for ion fluences above $1 \times 10^{14}$ cm$^{-2}$. The behavior of the (In,Fe)Sb layers is different.

TABLE 1. Experimental values of carrier concentration and mobility at RT for structure 250-0 irradiated with different fluences.

| Fluence, cm$^{-2}$ | Carrier concentration, cm$^{-3}$ | Mobility, cm$^2$/V·s |
|---|---|---|
| 0 | $7.5 \times 10^{17}$ ($n$-type) | 526 |
| $1 \times 10^{13}$ | $1.3 \times 10^{17}$ ($n$-type) | 430 |
| $1 \times 10^{14}$ | $2.2 \times 10^{18}$ ($p$-type) | 66 |
| $1 \times 10^{15}$ | $4.1 \times 10^{20}$ ($p$-type) | 3.5 |
| $1 \times 10^{16}$ | $6.0 \times 10^{20}$ ($p$-type) | 2 |

Figure 4 exhibits temperature dependences of the resistivity $\rho(T)$ for the as-grown samples 200-25 and 300-25 and for those irradiated with fluences of $1 \times 10^{14}$, $1 \times 10^{15}$ and $1 \times 10^{16}$ cm$^{-2}$. Before irradiation, the resistivity of sample 300-25 is much higher than that of sample 200-25. This is a consequence of a lower electron concentration resulting from a higher growth temperature (the concentration of native electrically active donor defects is lower). After irradiation the resistivity decreases in both structures. It is obvious that the resistivity decrease is related to a carrier density increase (the mobility should decrease after irradiation due to increasing number of scattering centres). The conductivity type conversion is observed (by Seebeck effect measurements) at fluences of $1 \times 10^{16}$ cm$^{-2}$ and $1 \times 10^{15}$ cm$^{-2}$ for samples 200-25 and 300-25, respectively.

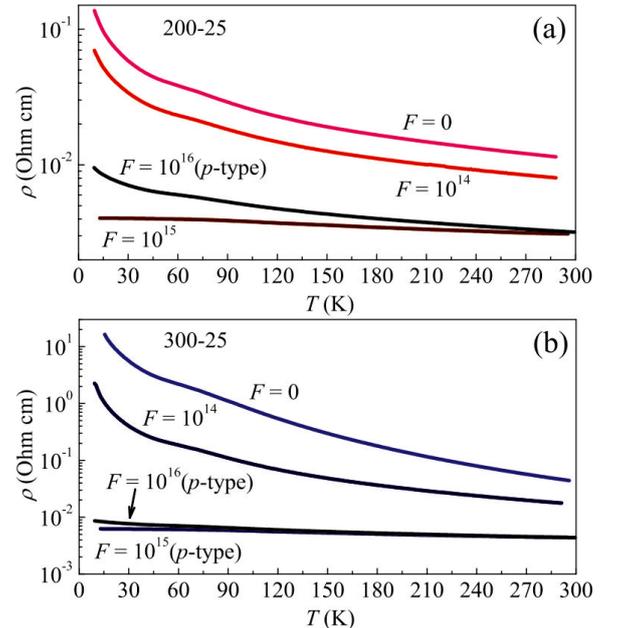

FIG. 4. Temperature dependences of the resistivity for structures 200-25 and 300-25 before and after irradiation.

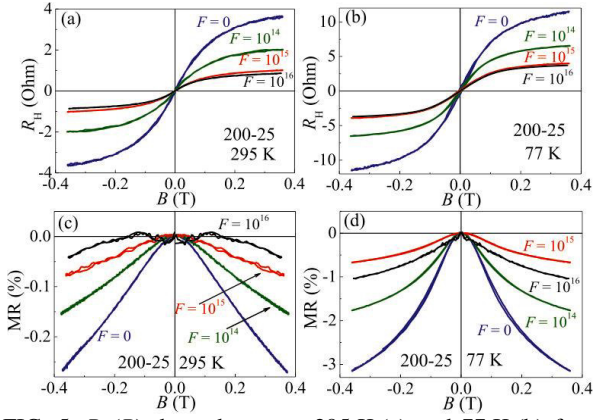
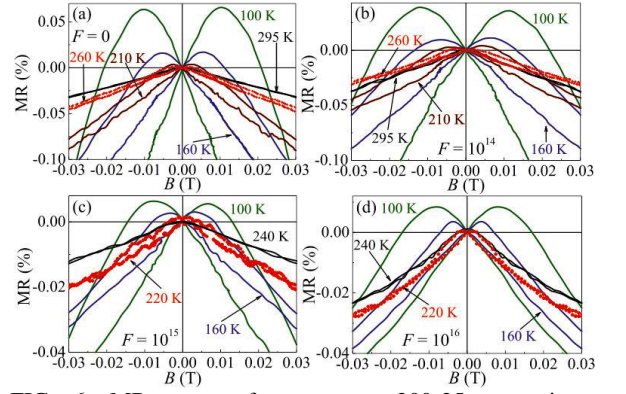

FIG. 5. $R_H(B)$ dependences at 295 K (a) and 77 K (b) for structure 200-25 before and after irradiation. Magnetoresistance curves at 295 K (c) and 77 K (d) for structure 200-25 before and after irradiation. $B$ is applied perpendicular to sample's surface.

Figure 5(a, b) shows Hall resistance dependences on the external magnetic field ($R_H(B)$) at 295 K and 77 K for sample 200-25 before and after irradiation. The $R_H(B)$ dependences are nonlinear with a saturation at $B \approx 0.2$ T, i.e. the anomalous Hall effect (AHE) is observed at both 77 K and RT. The $R_H(B)$ curves have the p-type sign both in the n-type and p-type state (the latter after irradiation with $F = 1 \times 10^{15}$ cm$^{-2}$) of sample 200-25. Consequently, the $R_H(B)$ dependences are completely determined by the AHE, and the ordinary Hall effect is not observed. Note that the shape of the $R_H(B)$ curves does not change with increasing carrier concentration and conductivity type conversion. The Hall resistance decrease is related to the resistivity decrease after irradiation. Figure 5(c, d) shows the magnetoresistance ($MR = (\rho(B) - \rho(0))/\rho(0)$) curves taken at 295 K and 77 K with the external magnetic field applied perpendicular to the (In,Fe)Sb layer for samplee 200-25 before and after irradiation. A negative MR is observed for both the as-grown and irradiated samples. The MR magnitude decreases with the ion fluence (Figure 5(c, d)). Perhaps this is related to some peculiarities of the negative MR in a system with two charge carrier types. The absence of the hysteresis in the $R_H(B)$ and MR curves for $B$ perpendicular to the layer is due to the predominant in-plane orientation of the easy magnetization axis.

Figure 6 exhibits the MR curves before and after irradiation for sample 200-25 in the temperature range from 100 – 295 K for $B$ applied in the plane of the structure and varying in the range of ± 0.03 T. In this case the MR curves are hysteretic. For the as-grown sample and that irradiated with a fluence of $1 \times 10^{14}$ cm$^{-2}$ the shapes of the MR curves are similar, and the clear hysteretic character of the in-plane MR curves is observed at 260 K (Figure 6(a, b)). Note that the weak hysteresis for the cases of $F = 0$ and $F = 1 \times 10^{14}$ cm$^{-2}$ presents on MR curves also at 295 K. Hence, $T_C$ for the as-grown sample 200-25 and that irradiated with $F = 1 \times 10^{14}$ cm$^{-2}$ is above RT. Further irradiation leads to a modification of the MR dependences.

FIG. 6. MR curves for structure 200-25 at various temperatures before and after irradiation. Magnetic field is applied in the sample plane.

Figure 7(a) shows the MR curves for sample 200-25 at temperatures between 240 – 260 K before and after irradiation (with $B$ applied in the sample plane). For $F = 0$ and $F = 1 \times 10^{14}$ cm$^{-2}$ the dependences are hysteretic at 260 K, while for $F = 1 \times 10^{15}$ and $1 \times 10^{16}$ cm$^{-2}$ the hysteresis disappears at 240 K.

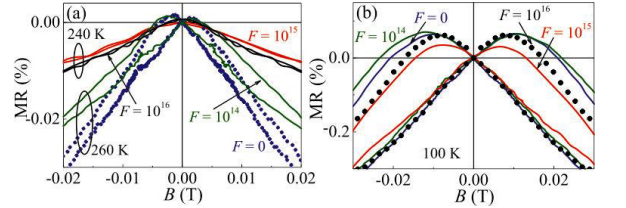

FIG. 7. (a) MR curves for structure 200-25 at 240 - 260 K before and after irradiation. (b) MR curves for structure 200-25 at 100 K before and after irradiation. Magnetic field is applied in the sample plane.

Figure 7(b) shows a comparison of the MR curves at 100 K (with $B$ applied in the sample plane) for different $F$ values (the curve for $F = 1 \times 10^{14}$ cm$^{-2}$ is multiplied by a factor of 1.8, the curves for $F = 1 \times 10^{15}$ and $1 \times 10^{16}$ cm$^{-2}$ are multiplied by a factor of 5.5). For $F = 1 \times 10^{15}$ and $1 \times 10^{16}$ cm$^{-2}$ a shift of the positive MR peaks to lower magnetic fields is observed, which indicates a coercivity decrease. The evolution of the shape of the MR curves with the ion fluence also indicates some weakening of the ferromagnetic properties of sample 200-25 after irradiation with $F = 1 \times 10^{15}$ and $1 \times 10^{16}$ cm$^{-2}$.

The magnetoresistance studies for sample 200-25 are consistent with MFM studies. Figure 8 shows the MFM and corresponding AFM images obtained at RT for as-grown sample 200-25 and after irradiation with a fluence of $1 \times 10^{16}$ cm$^{-2}$. The MFM image for as-grown structure has a weak but well detectable magnetic contrast (Figure 8 (a)), which clearly differs from the surface morphology of the same part of the structure (Figure 8 (b)). Hence, the MFM studies confirm that $T_C$ for the as-grown sample 200-25 is above RT. The irradiation with $F = 1 \times 10^{16}$ cm$^{-2}$ leads to a noticeable weakening of the magnetic contrast (Figure 8(c)). This is consistent with the magnetoresistance results about the weakening of the ferromagnetic properties of sample 200-25 after irradiation with $F = 1 \times 10^{15}$ and $1 \times 10^{16}$ cm$^{-2}$.

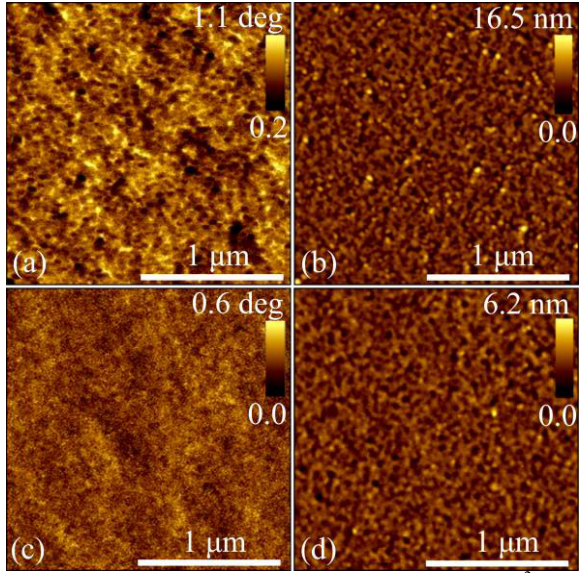

FIG. 8. MFM (a) and AFM (b) images of $1 \times 1$ μm$^2$ area for as-grown structure 200-25. MFM (c) and AFM (d) images of $1 \times 1$ μm$^2$ area for structure 200-25 after irradiation with $F = 1 \times 10^{16}$ cm$^{-2}$.

Note that the very weak but detectable MFM contrast at RT is also observed after irradiation (Figure 8(c)). The surface studies by Kelvin probe force microscopy before and after irradiation reveal that the distribution of the surface potential does not coincide with the obtained magnetic contrast, which confirms the correctness of the MFM results.

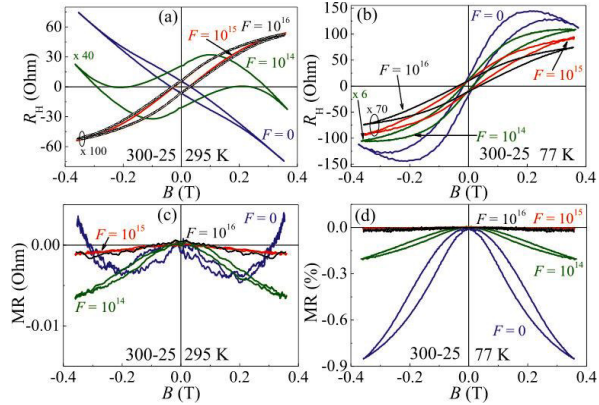

FIG. 9. $R_H(B)$ dependences at 295 K (a) and 77 K (b) for structure 300-25 before and after irradiation. MR curves at 295 K (c) and 77 K (d) for structure 300-25 before and after irradiation. Magnetic field is applied perpendicular to the sample plane.

Figure 9(a, b) shows the $R_H(B)$ dependences at 295 K an 77 K for sample 300-25 before and after irradiation. Unlike for sample 200-25, the $R_H(B)$ curves for sample 300-25 have a pronounced hysteresis at RT and 77 K, and their shape significantly changes after irradiation. However, the hysteretic shape of the $R_H(B)$ curves is not related to the true anomalous Hall effect. This is the ordinary Hall effect (OHE) in the (In,Fe)Sb conductive layer with Fe ferromagnetic inclusions. We observed a similar anomalous-like OHE in (In,Mn)As layers with MnAs clusters, and it was explained by the Lorentz force caused by the magnetic field of ferromagnetic MnAs inclusions and by an inhomogeneous distribution of the current density in the layer [20]. The irradiation leads to a significant carrier density increase, which affects the magnitude of the Hall resistance and the shape of the $R_H(B)$ curves, as we considered in detail in Ref. [20]. The hysteretic dependence of the average magnetization of ferromagnetic inclusions (and therefore the effective internal magnetic field in the layer) on the external magnetic field results in hysteretic MR dependences (Figure 9(c, d)).

The linear part of the $R_H(B)$ curve at 295 K for the as-grown sample 300-25 (Figure 9(a)) allows us to determine the electron concentration of $3 \times 10^{17}$ cm$^{-3}$. In view of the lower resistance of sample 200-25 at RT (Figure 4), the electron concentration in the as-grown sample 200-25 can be estimated to be above $1 \times 10^{18}$ cm$^{-3}$. After irradiation with $F = 1 \times 10^{15}$ cm$^{-2}$, sample 200-25 remains $n$-type, and the electron concentration can be estimated (taking into account the resistivity decrease (Figure 4(a)) and the mobility decrease) to be about $1 \times 10^{19}$ cm$^{-3}$.

## IV. DISCUSSION

Let us discuss the influence of ion irradiation on the single-phase DMS (In,Fe)Sb layer of sample 200-25. The results of previous experimental studies indicate that the ferromagnetism and high Curie temperature in (In,Fe)Sb are not directly related to the charge carrier concentration [4,5]. In Ref. [21] dedicated to the electric field effect in the (In$_{0.89}$,Fe$_{0.11}$)Sb layer, $T_C$ was varied between 207 – 216 K with the electron density variation in the range from $3.6 \times 10^{17}$ – $7.5 \times 10^{17}$ cm$^{-3}$, and it was suggested that the electron induced mechanism of the ferromagnetism coexists with some other predominant mechanism (probably superexchange). In the theoretical work [22] it was concluded that the superexchange mechanism in (In,Fe)Sb and (Ga,Fe)Sb produces antiferromagnetic interactions between isoelectronic Fe atoms, and the ferromagnetic interactions should appear due to the double exchange appearing after the Fermi level shift into the conduction or valence bands (as a result of the $n$- or $p$-type doping). It was assumed that $T_C$ depends on the Fermi level position. However, the double exchange mechanism requires free carriers to provide the exchange interaction. The estimation of the relationship between $T_C$ and the carrier concentration was not carried out in Ref. [22]. The irradiation of sample 200-25 allows to change both the carrier concentration and the Fermi level position. As mentioned above, the irradiation with a fluence of $1 \times 10^{15}$ cm$^{-2}$ results in a significant increase in the electron concentration (about an order of magnitude) in sample 200-25. This electron concentration increase does not lead to an increase of $T_C$ or coercivity. Inversely, a weakening of the ferromagnetic properties after irradiation with a fluence of $1 \times 10^{15}$ cm$^{-2}$ was observed (Figures 6 and 7). Note that after the conductivity type conversion (at $F = 1 \times 10^{16}$ cm$^{-2}$) no further noticeable changes in the magnetic properties were revealed (Figure 7). For $F = 0$, $1 \times 10^{14}$ and $1 \times 10^{15}$ cm$^{-2}$, sample 200-25 continues $n$-type. In these cases the concentration of electrically active donor defects exceeds that of electrically active acceptor defects, and the resulting electron density is quite high (~ $10^{18} - 10^{19}$ cm$^{-3}$), consequently, the Fermi level is located in the conduction band. Figure 10(a) shows a band diagram of the $n$-InSb/$i$-GaAs structure at 77 K calculated with Gregory Snider's 1D Poisson/Schrödinger solver [23]. For the modeling of the $n$-type InSb, the ionized donors concentration $N_D = 2.1 \times 10^{19}$ cm$^{-3}$ and the ionized acceptors concentration $N_A = 2.0 \times 10^{19}$ cm$^{-3}$ were taken. After the conductivity type conversion (at $F = 1 \times 10^{16}$ cm$^{-2}$) the concentration of electrically active acceptor defects exceeds that of electrically active donor defects, and the resulting hole density is also quite high (~ $10^{19}$ cm$^{-3}$, since the resistivity remains low (Figure 4(a))). Consequently, the Fermi level shifts into the valence band. Figure 10(b) shows the calculated band diagram of $p$-InSb/$i$-GaAs structure at 77 K ($N_A = 2.1 \times 10^{20}$ cm$^{-3}$, $N_D = 2.0 \times 10^{20}$ cm$^{-3}$).

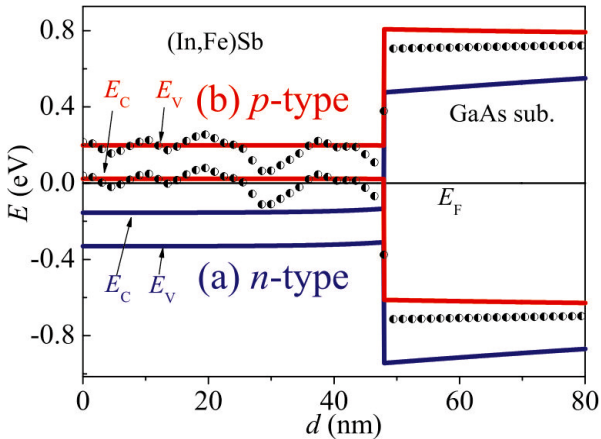

FIG. 10. Calculated band diagrams for the InSb/$i$-GaAs structure at 77 K. Solid lines demonstrate the $n$-type case (before conductivity type conversion, $N_D = 2.1 \times 10^{19}$ cm$^{-3}$, $N_A = 2.0 \times 10^{19}$ cm$^{-3}$) and $p$-type case (after conductivity type conversion, $N_D = 2.0 \times 10^{20}$ cm$^{-3}$, $N_A = 2.1 \times 10^{20}$ cm$^{-3}$). Semi-filled circles illustrate the case of spatial fluctuation of acceptors and donors with the predominance of acceptors ($N_A = 2.0 \times 10^{20}$ cm$^{-3}$, $N_D = (1.8 – 2.2) \times 10^{20}$ cm$^{-3}$).

As noted above, the irradiation of sample 200-25 with fluences of $1 \times 10^{15}$ and $1 \times 10^{16}$ cm$^{-2}$ leads to the weakening of the ferromagnetic properties. This result is basically consistent with the predicted dependence of magnetic properties on the Fermi level position for (In,Fe)Sb [22]. However, the observed changes in the magnetic properties are not drastic. In principle, the observed weakening of the ferromagnetic properties after irradiation can have a different origin. The ion irradiation creates in (In,Fe)Sb a random spatial distribution of electrically active donor and acceptor defects. In particular, after the conductivity type conversion, although the concentration of acceptor centers predominates, there are local areas with a different compensation degree. The semi-filled circles in Figure 10(b) illustrate the band edges in case of a non-uniform distribution of electrically active defects. This leads to strong spatial fluctuations of the built-in electric field, which can weaken the superexchange interaction between Fe atoms by modifying the electron density distribution around the intermediate non-magnetic atoms.

## V. CONCLUSION

The influence of 50 keV He$^+$ ion irradiation on the transport and magnetic properties of a nominally undoped InSb layer, a single-phase (In,Fe)Sb DMS layer and a two-phase (In,Fe)Sb layer with Fe inclusions was investigated. The initially $n$-type InSb layer demonstrates a conductivity type conversion after irradiation with a fluence of $1 \times 10^{14}$ cm$^{-2}$. The irradiation of the (In,Fe)Sb layers reveals the formation of both the acceptor- and donor-type electrically active RDs. The $n$- to $p$-type conversion was observed in In$_{1-x}$Fe$_x$Sb matrices with $x \sim 0.03$ and $x = 0.26$ after irradiation with fluences of $1 \times 10^{15}$ and $1 \times 10^{16}$ cm$^{-2}$, respectively. The magnetotransport properties of the two-phase (In,Fe)Sb layer are strongly affected by ferromagnetic Fe inclusions. An influence of the density of electrically active RDs on the magnetic properties of the single-phase DMS (In,Fe)Sb was found. The observed increase of the majority carrier (electrons) concentration by about an order of magnitude after irradiation with a fluence of $1 \times 10^{15}$ cm$^{-2}$ is accompanied by a weakening of the ferromagnetic properties. The change in the type of the majority carriers (from electrons to holes) after irradiation with a fluence of $1 \times 10^{16}$ cm$^{-2}$ does not lead to further changes in the magnetic properties. In general, we can conclude that the magnetic properties of the DMS (In,Fe)Sb are quite resistant to significant changes in the charge carrier concentration and the Fermi level position. The results confirm a weak interrelation between the ferromagnetism and the charge carrier concentration in (In,Fe)Sb.

## ACKNOWLEDGMENT

This study was supported by Russian Science Foundation (grant № 18-79-10088). N.A.S. gratefully acknowledges the support of the Ministry of Education and Science of the Russian Federation in the framework of the Increase Competitiveness Program of NUST «MISiS» (no. K3-2018-025), implemented by a governmental decree dated 16th of March 2013, no. 211, and the support of the FCT of Portugal through the Project No. I3N/ FSCOSD (Ref. FCT UID/CTM/50025/2013).